\documentclass[preprint]{aastex61}
\usepackage{amsmath}
\usepackage{graphicx}
\usepackage[outdir=./]{epstopdf}
\usepackage{natbib}
\shorttitle{On the ratios of two \ion{Si}{4} lines}
\shortauthors{Tripathi et al.}

\begin{document}
\title{On the ratios of \ion{Si}{4} lines ($\lambda$1394/$\lambda$1403) in an emerging flux region}
\correspondingauthor{Durgesh Tripathi }
\email{durgesh@iucaa.in}
\author[0000-0003-1689-6254]{Durgesh Tripathi}
\affil{Inter-University Centre for Astronomy and Astrophysics, Post Bag {--} 4, Ganeshkhind, Pune 411007, India}
\author{Nived V N}
\affil{Inter-University Centre for Astronomy and Astrophysics, Post Bag {--} 4, Ganeshkhind, Pune 411007, India}
\affil{Armagh Observatory and Planetarium, College Hill, Armagh BT61 9DG, UK}
\author[0000-0003-2976-5130]{Hiroaki Isobe}
\affil{Faculty of Fine Arts, Kyoto City University of Arts, 13-6 Kutsukake-cho, Oe, Nishikyo-ku, Kyoto, Kyoto 610-1197, Japan}
\author{J. Gerry Doyle}
\affil{Armagh Observatory and Planetarium, College Hill, Armagh BT61 9DG, UK}

\begin{abstract} 
The resonance lines of \ion{Si}{4} formed at $\lambda$1394 and 1403 {\AA} are the most critical for the diagnostics of the solar transition region in the observations of the Interface Region Imaging Spectrograph (IRIS). Studying the intensity ratios of these lines (1394{\AA}/1403{\AA}), which under optically thin condition is predicted to be two, helps us to diagnose the optical thickness of the plasma being observed. Here we study the evolution of the distribution of intensity ratios in 31 IRIS rasters recorded for four days during the emergence of an active region. We found that during the early phase of the development, the majority of the pixels show intensity ratios smaller than two. However, as the active region evolves, more and more pixels show the ratios closer to two. Besides, there are a substantial number of pixels with ratio values larger than 2. At the evolved stage of the active region, the pixels with ratios smaller than two were located on the periphery, whereas those with values larger than 2 were in the core. However, for quiet Sun regions, the obtained intensity ratios were close to two irrespective of the location on the disk. Our findings suggest that the \ion{Si}{4} lines observed in active regions are affected by the opacity during the early phase of the flux emergence. The results obtained here could have important implications for the modelling of the solar atmosphere, including the initial stage of the emergence of an active region as well as quiet Sun.
\end{abstract}

\section{INTRODUCTION} \label{sec:intro}

The transition region is the atmospheric layer that separates the cooler chromosphere from the hotter corona. Studying the structure and dynamics of the transition region requires greater attention as these provide critical information on the supply of mass and energy from the lower atmosphere to the corona and the solar wind. Emission lines originating in the transition region mostly fall in the ultraviolet part of the spectra. These UV photons can propagate through the upper atmosphere without significant absorption, re-emission and scattering. Thus transition region emission lines are generally considered as optically thin. However, some of the transition region lines are affected by opacity effects. For example emission lines of \ion{Si}{4}, \ion{C}{4}, \ion{Si}{3}, \ion{C}{3}, \ion{Si}{2}, \ion{C}{2} show the effects of optical thickness \citep{keenan,doyle,brook,mason}.

How do we determine if the emission line understudy is optically thin or thick? This problem have been studied in great detail by \cite{jordan}, \cite{doyle}, \cite{keenan}. According to \cite{jordan}, under the optically thin condition, the ratio of intensities of two lines originating from the same upper level is simply the ratio of their transition probabilities. For such lines, any deviation from this theoretical ratio value would be the effects of opacity. For Li-like (e.g., \ion{C}{4}) resonance lines, this ratio is exactly two \citep{yan}. Similarly, as theoretically predicted \citep[e.g. by CHIANTI;][]{chianti_15,chianti_1}, in optically thin conditions, the intensity of the \ion{Si}{4}~1394~{\AA} line is twice the intensity of the \ion{Si}{4}~1403~{\AA} line, though we note that the two \ion{Si}{4} lines originate from two different levels, namely 2P3/2 and 2P1/2, respectively. However, in optically thick conditions, the intensity of the stronger line decreases and the ratio of these lines can be less than two and in some instances greater than two (see \cite{keenan_2014} work on the Li-like ion O VI). The intensity of the stronger line changes because the line with the largest oscillator strength is strongly affected by opacity \citep{mason,keenan_2014}. In other words, the deviation of the ratio of the resonance lines from its theoretical value provides potentially useful information about the physical environment of the transition region and the atmosphere above. For example, \cite{yan} confirmed the presence of self-absorption features using \ion{Si}{4} line ratios. During extreme ultraviolet (EUV) brightening event, they found a decrease in \ion{Si}{4} ratio from 2 to 1.7. It was suggested that the self-absorption is either due to the pre-existing \ion{Si}{4} ions in the upper atmosphere or the self-absorption is arising within the region of brightness. Similarly, \cite{resonant_sct} observed pixels with intensity ratios greater than 2 in an active region and suggested that they are due to resonant scattering.

The main aim of this paper is to study the opacity effects on the two \ion{Si}{4} lines in an emerging flux region (EFR) and how these effects vary during the evolution of the active region. Solar magnetic flux is generated by the dynamo process in the tachocline -- the region that separates the radiative zone from the convection zone. Due to magnetic buoyancy \citep{parker}, the flux then rises and emerges through the photosphere forming active regions and sunspots. These are called emerging flux regions \citep{zirin}. Interaction of this emerging flux with pre-existing coronal magnetic field gives rise to a wide variety of phenomena such as X-ray jets, explosive events, flares, coronal mass ejections (CMEs) etc. \citep[see, e.g.][]{chifor2,chifor1}. In H$\alpha$ EFRs show the presence of both bright and dark loops \citep[]{zirin_filaments,bruzek_filaments}. The dark and thick threads of loops are known as arch filaments; they are considered as the trace of rising flux tubes \citep{chou}. The rising magnetic flux lifts cool and dense chromospheric plasma into the corona, forming so-called arch filament system \cite{isobe}. Due to the intermittent nature of heating, hot and cool plasmas often coexist in EFRs \citep{isobe06}. Moreover, below the emerging loops there usually exist small scale brightenings in the chromosphere, called Ellerman bombs, produced by magnetic reconnection between the newly emerging magnetic loops \citep[][]{pariat, IsoTA}. Recent observations from IRIS found that similar "bombs" also seen as UV brightenings \citep[][]{peter, GupT}. Therefore, EFR is a likely place that will have large densities along the line of sight, where the opacity effects may play a significant role in the formation of transition region lines in EFRs.

For the above-described purpose, we use the observations recorded by the Interface Region Imaging Spectrograph (IRIS, \cite{iris}) observations. The very high spatial and spectral resolution of IRIS provides us with an excellent opportunity to study the transition region in great detail. The rest of the paper is structured as follows. We provide the details of observation and data analysis in \S\ref{sec:od}. The obtained results are presented in \S\ref{sec:results}. Finally, we discuss the results and conclude in \S\ref{sec:conclusion}.

\begin{figure}[h]
 \centering
 \includegraphics[width=0.8\textwidth]{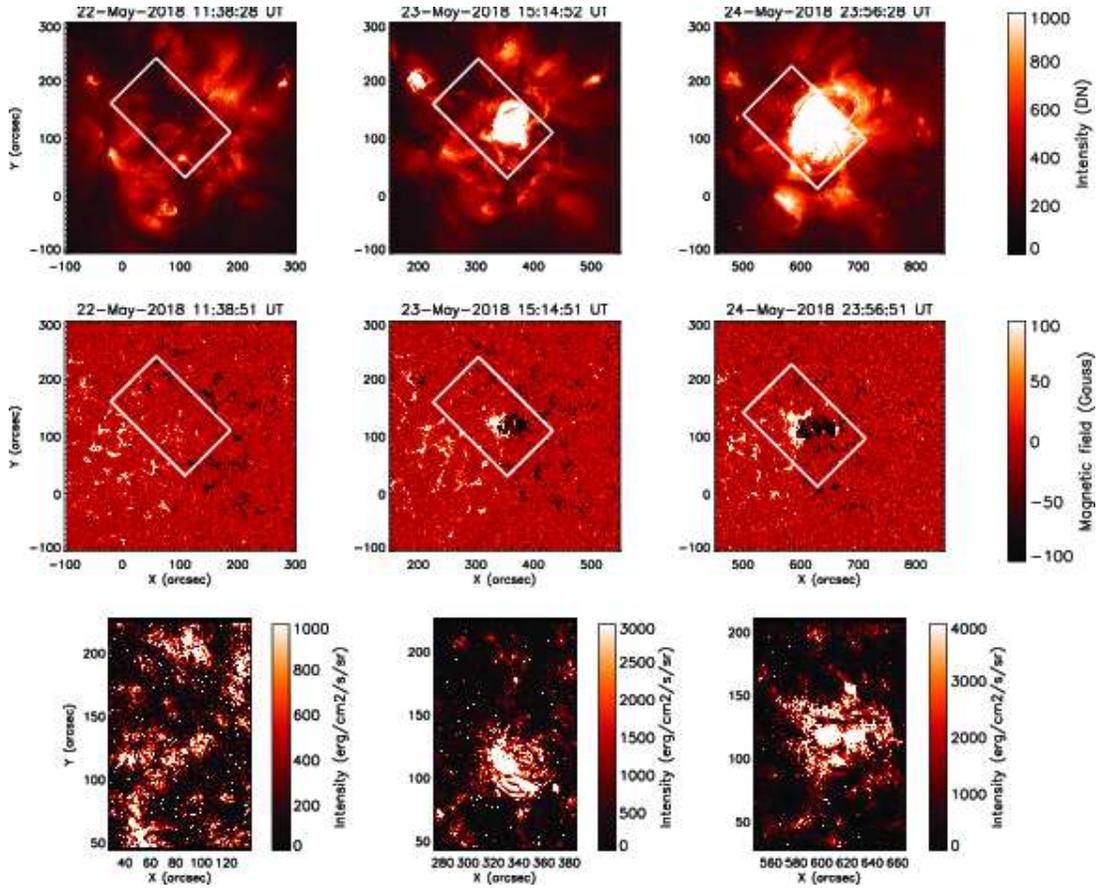}
 \caption{ Top panels: A portion of Sun's disk observed by AIA using 193 {\AA} channel during the evolution of the active region. Middle panels: HMI magnetograms corresponding to the FOV of coronal images shown in the top rows. The over-plotted rectangles display the IRIS raster field of view. Bottom panels: IRIS raster images obtained in \ion{Si}{4}~1394~{\AA}.}
 \label{fig:fig1}
\end{figure}

\section{Observations and Data Analysis}\label{sec:od}

The primary aim of this work is to study the ratios and the evolution of the two \ion{Si}{4} lines formed at 1394~{\AA} and 1403 {\AA} observed by IRIS in an emerging flux region. For this purpose, we have analysed 31 IRIS rasters obtained over emerging active region NOAA 12711 that eventually evolves into an old active region. IRIS obtains UV spectra in far UV (1332 to 1407 {\AA}) and in near UV (2783 to 2835 {\AA}) with an effective spectral resolution of 26 m{\AA} and 53 m{\AA} respectively \citep{iris}. The effective spatial resolution of the far UV spectra is 0.33{\arcsec} and that for near UV spectra is 0.4{\arcsec}. The rasters were recorded during 11:24 UT on 2018 May 22 to 02:00 UT on 2018 May 25. In this study, we have also used the coronal observations recorded with the Atmospheric Imaging Assembly \citep[AIA][]{aia, aia2, aia3} and line-of-sight (LOS) magnetic field maps obtained by the Helioseismic magnetic imager \citep[HMI;][]{hmi}, both on board the Solar Dynamics Observatory.

Fig.~\ref{fig:fig1} displays three snapshots of the emerging active region as was observed with the AIA in 193~{\AA} passband (top panels) and HMI (middle panels). The over-plotted white boxes in the top and middle panels show the IRIS raster FOV. The corresponding \ion{Si}{4} intensity maps are shown in the bottom panels. Note that we have applied standard software for data processing and analysis provided in \textsl{SolarSoft} \citep{solarsoft}.

All IRIS observations were taken in dense raster mode. The exposure time for most of the rasters was 4~s. However, exposure times of eight and 15~s were used for a few rasters. We note that since we are interested in studying the ratio of the lines, the varying exposure times do not affect our results. We provide the details of the rasters in Table~\ref{table:tab1} with their corresponding $\mu$ values, where $\mu$ is defined as the cosine of the longitude. Thus, $\mu$ values close to 1 represent the disk centre observations, whereas those close to 0 represent the limb observations. 

IRIS provides intensity values in the units of counts (DN). To convert IRIS spectra to a physical unit (ergs~cm$^{-2}$~s$^{-1}$~sr$^{-1}$), we performed radiometric calibration\footnote{IRIS technical note 24}. We then obtained the intensity maps for \ion{Si}{4} 1394 and 1403 {\AA} line by fitting a Gaussian to every pixel. The ratio maps were subsequently generated by dividing the 1394 {\AA} maps with 1403 {\AA} maps. As mentioned earlier, ideally, under the optically thin condition, the intensity ratios should be precisely 2. However, we found that the ratio lied within a large range of values around 2. This obtained deviation from the theoretical line ratio could be attributed to the opacity effects as well as the bad fitting of the spectra. Therefore, it is necessary to remove all the poorly fitted and noisy data before studying the statistics of ratio values and their evolution. 

\begin{figure}
 \centering
 \includegraphics[width=0.8\textwidth]{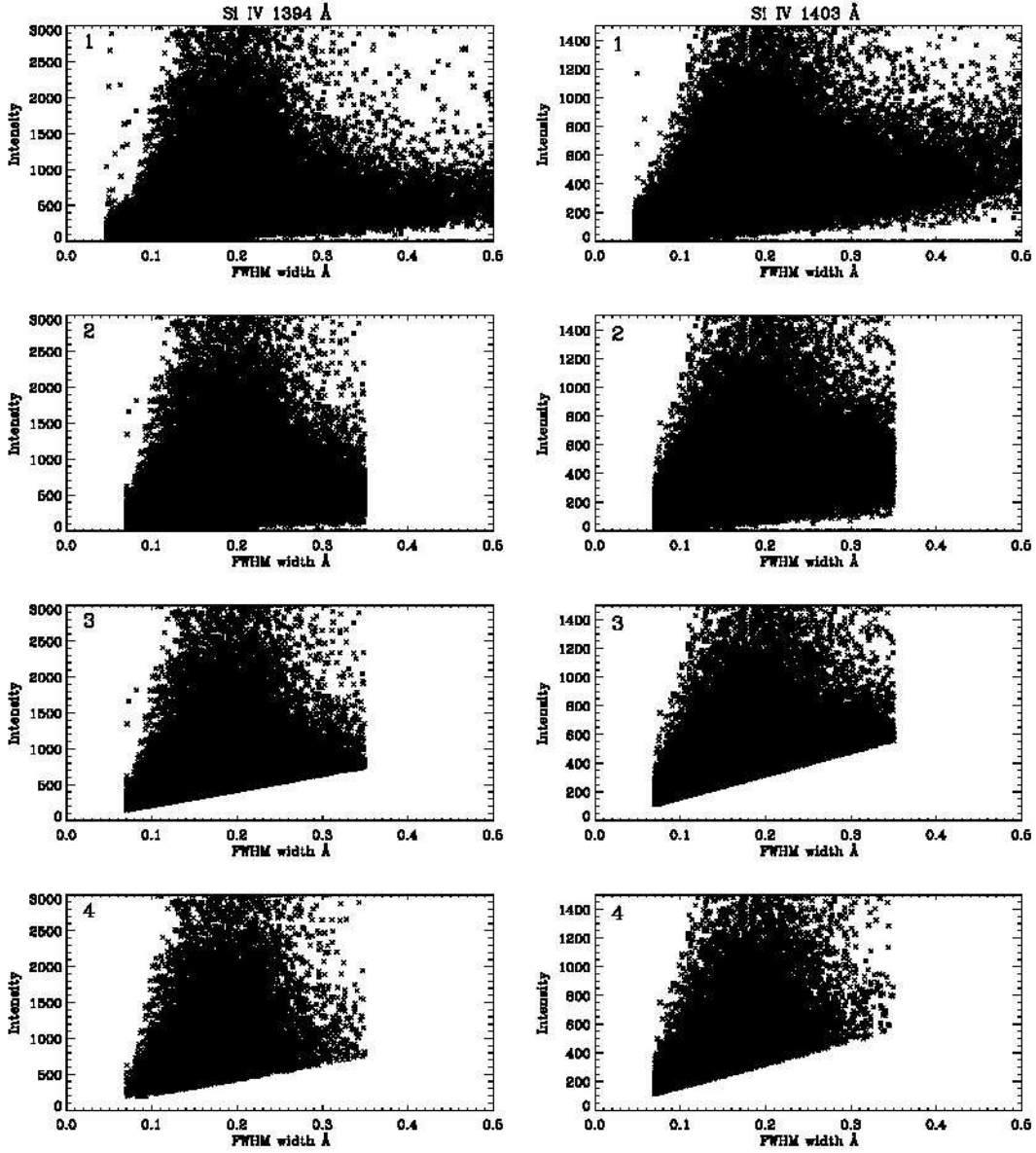}
 \caption{Scatter plot of intensity as a function of full width at half maxima for 1394 {\AA} (left column) and 1403 {\AA} (right column) after applying the filtering condition given in \S\ref{sec:od}, step wise. Numbers on the top left corner of each plot correspond to the filtering process applied.}
 \label{fig:fig3}
\end{figure}

To get away from poorly fitted and noisy spectra, which could potentially affect the intensities and thereby the ratios, we apply the following filtering conditions (similar to \citet{resonant_sct}).

\begin{enumerate}
\item We applied the Gaussian fitting at each pixel within the raster.
\item We consider only those pixels for which the full-width half maximum (FWHM) of the fitted Gaussian lies within 70 m{\AA} and 350 m{\AA} in both the spectral lines of \ion{Si}{4}. The purpose of this condition is to identify badly fitted spectra. Moreover, pixels with cosmic ray hit, which could not be removed by de-spiking, could also lead to incorrect Gaussian fitting. Such pixels are also removed using this condition.
\item In addition, for each raster, we demanded that the peak of the fitted Gaussian should be more than 50 ergs~cm$^{-2}$~s$^{-1}$~sr$^{-1}$ for 1394~{\AA} and 38 for 1403~{\AA}. Note that we found this threshold values by trial and error method. This additional condition takes care of the dark region in the raster where the signals are very weak.
\item Finally, we perform our statistics only on those pixels that showed ratios between 1.5 and 2.5. Other higher and lower ratios may be unphysical and therefore, removed from our analysis. 
\end{enumerate}
 
The results of filtering processes on IRIS raster recorded on 2018 May 22 at 11:24:00 UT after each step are shown in Figure \ref{fig:fig3}. The scatter plot of intensity in 1394 and 1403 {\AA} line as a function FWHM are displayed in the left and right panels, respectively. The numbers mentioned in the left top corner of each plot corresponds to the filtering process applied. The scatter plots in the first row are plotted after removing pixels which are polluted by cosmic rays. The scatter plots shown in the fourth row are obtained by applying all four filtering conditions, as mentioned earlier. We continue our analysis on this data. We note that by using such criteria, we may have missed some of the genuine pixels. However, this does not significantly affect the statistical result.

\section{Results}\label{sec:results}
\subsection{Intensity Ratio Maps}
\begin{figure}[h]
 \centering
 \includegraphics[width=\textwidth]{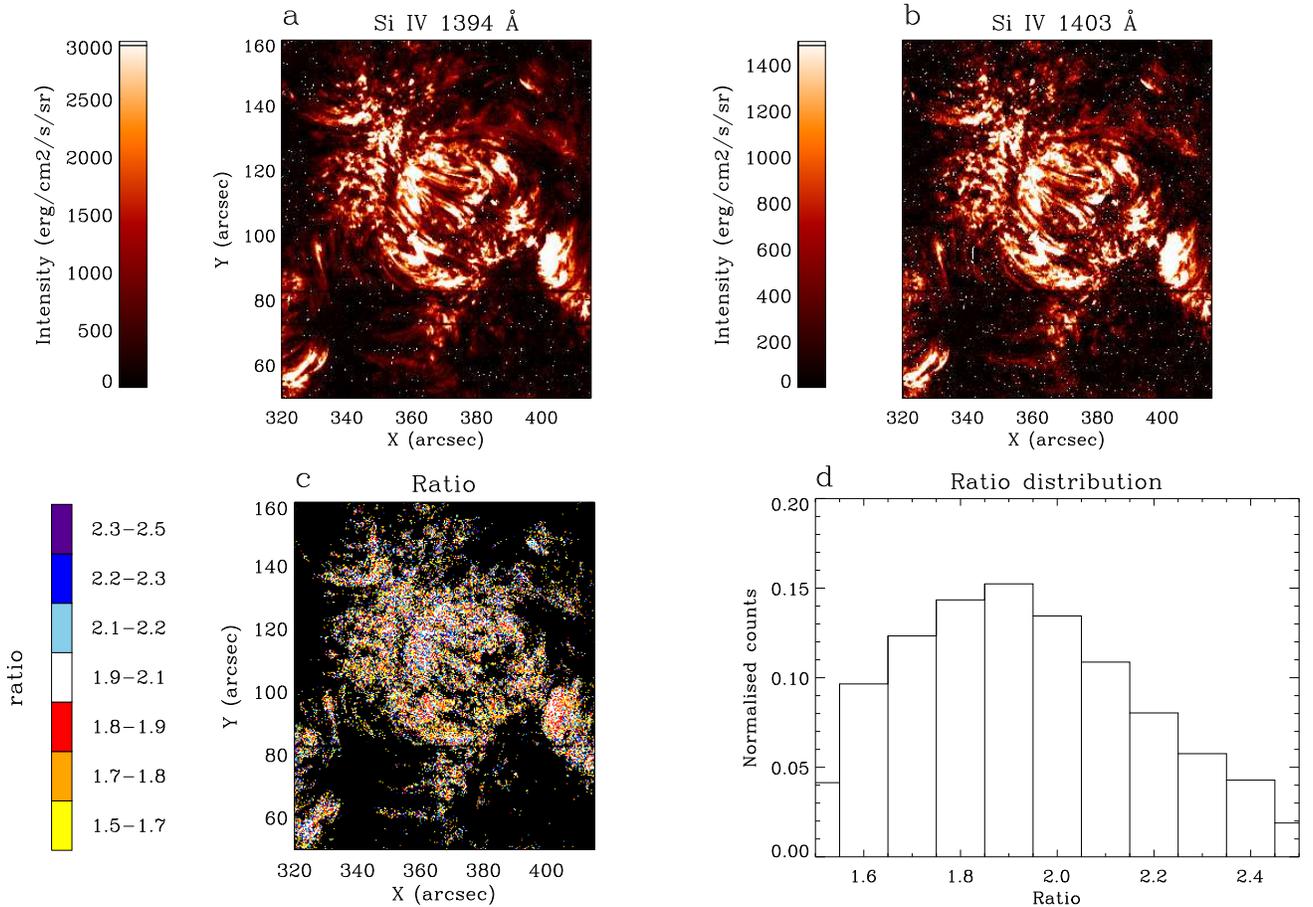}
 \caption{IRIS raster images obtained in \ion{Si}{4}~1394~{\AA} (panel a) and 1403~{\AA} (panel b) on 2018 may 23 18:28 UT. The obtained ratio map and the distribution of the ratio is shown in panels c and d, respectively.} \label{fig:fig4}
\end{figure}
 
 Figure~\ref{fig:fig4} presents the intensity maps obtained on May 23 18:28 UT for the \ion{Si}{4}~1394~{\AA} (top left), \& 1403~{\AA} (top right). The bottom left panel shows the obtained ratio map. The white coloured pixels in the ratio map represent the line ratio value between 1.9 and 2.1. The pixels with black colour are the missing pixels. The red, orange, yellow coloured pixels correspond to the ratio values less than 1.9 and the blue, violet, sky blue colours represent the pixels with a ratio above 2.1 as shown in the colour bar. From the ratio map, it is clear that the low ratio values are located at the periphery of the active region and a large fraction of the active region has a ratio value between 1.9 and 2.1. We plot the histogram of the corresponding ratio map in the bottom right panel. The y-axis of the histogram corresponds to the fraction of pixels which falls in each bin. As can be seen, the peak of the histogram is at 1.9. However, there are a significant number of pixels with ratios smaller as well as larger than 2. Note that this raster was taken when the active region was going through the evolutionary process.

\begin{figure}[h]
 \centering
 \includegraphics[width=\textwidth]{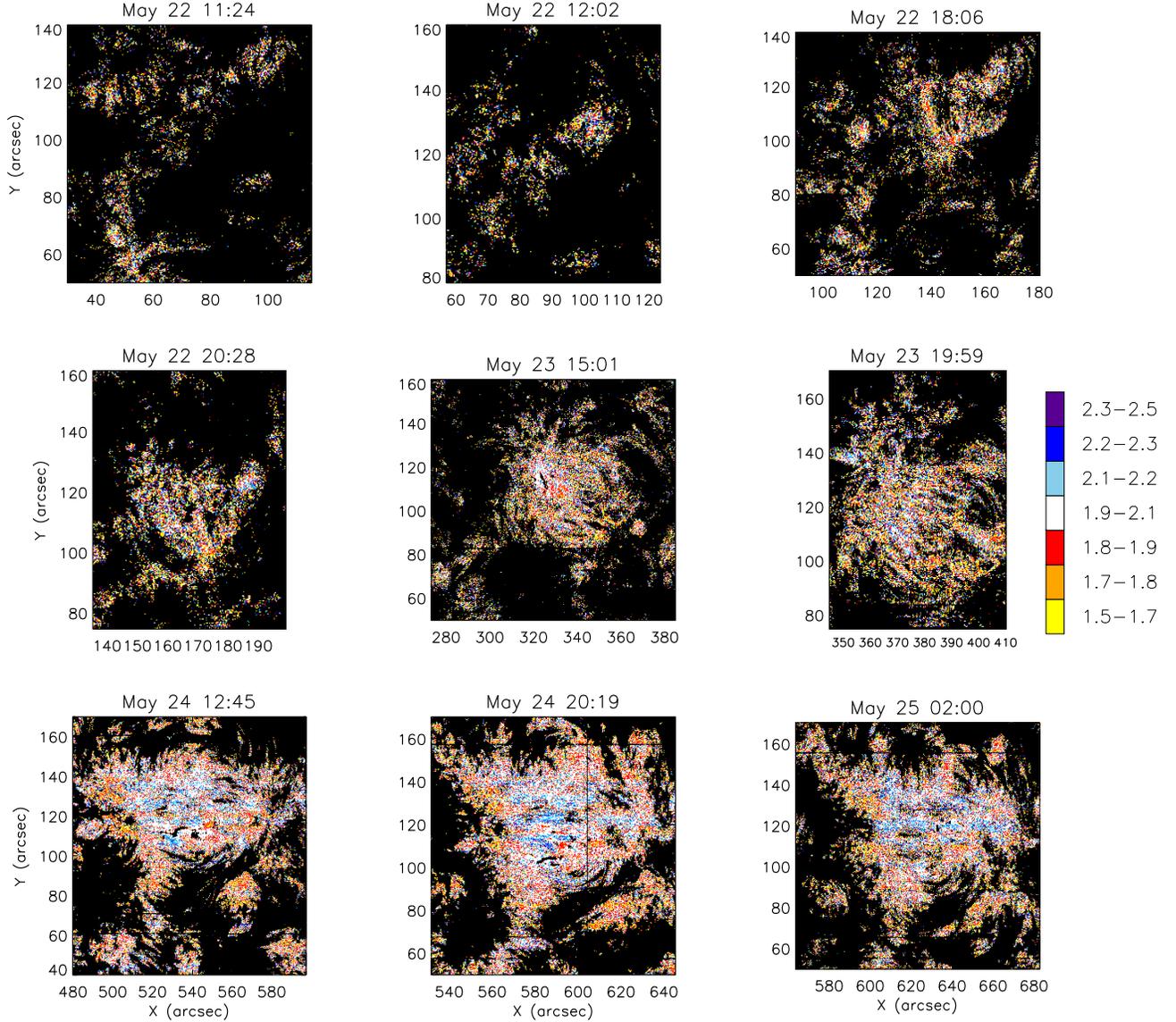}
 \caption{Evolution of the \ion{Si}{4} intensity ratio maps. The date and time is labelled.}
 \label{fig:fig5}
\end{figure}

\begin{figure}[h]
 \centering
 \includegraphics[width=0.8\textwidth]{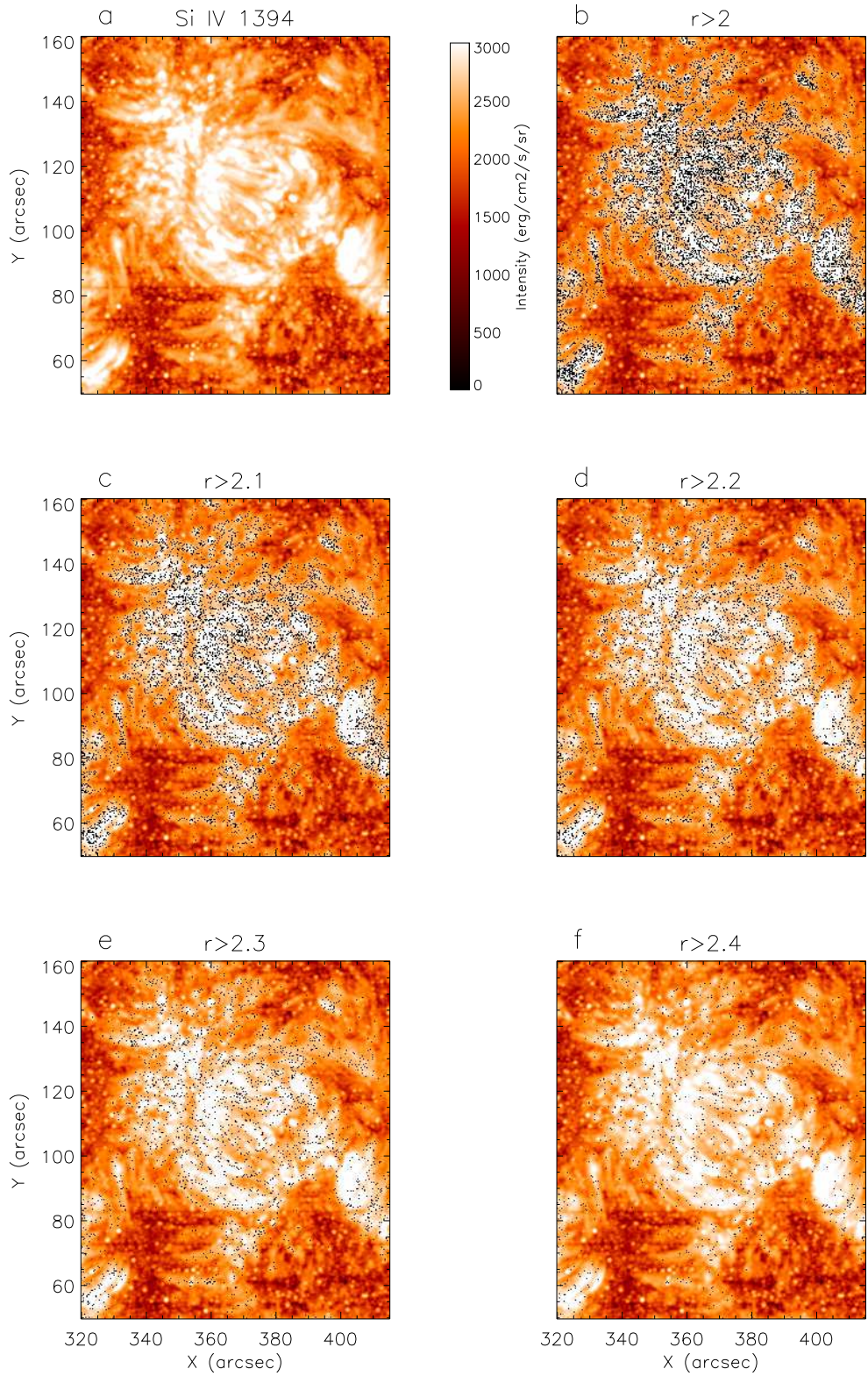}
 \caption{Panel a: \ion{Si}{4} 1394 {\AA} intensity image recorded on 23 May 2018 18:28 UT. Location of pixels with line ratio above 2 (panel:b), 2.1 (panel:c), 2.2 (panel:d), 2.3 (panel:e), 2.4 (panel:f) }
 \label{fig:fig6}
\end{figure}

\begin{figure}[h]
 \centering
 \includegraphics[width=0.8\textwidth]{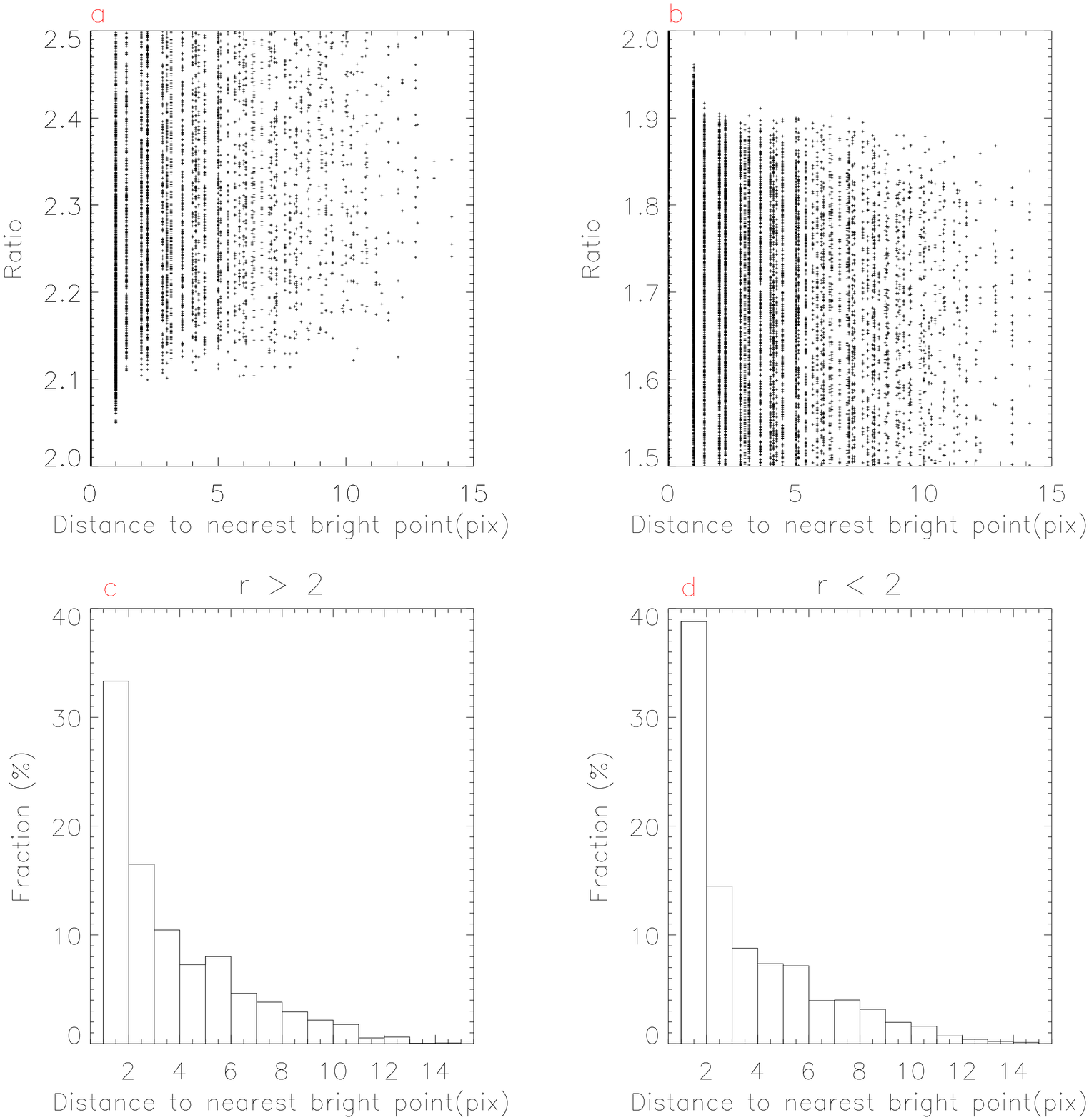}
 \caption{Scatter plot for intensity ratio (Panel a: for ratios $>$2; Panel b: ratios $<$ 2) as function of distance to nearest bright point. The corresponding histograms are shown in panel c and d, respectively.}
 \label{fig:distance}
\end{figure}

To study the temporal evolution of the intensity ratios, in Fig.~\ref{fig:fig5}, we plot the intensity ratio maps of the active region. The dates and time (in UT) of the rasters are labelled. The observation recorded on 22nd May represent the initial stage of flux emergence, where there are no bright structures observed as yet in AIA 193~{\AA} images (see Fig.~\ref{fig:fig1}). Whereas those taken on May 23, 24, 25 show the development of bright structures with time, thus they represent the growth phase of the active region. In the ratio maps, all the pixels with ratios lying between 1.9 and 2.1 are shown in white (see the colour bar in Fig.~\ref{fig:fig5}). As can be depicted from the ratio maps, during the initial phase of the emergence, the ratio maps contain a significant fraction of pixels with ratios different from being between 1.9 and 2.1. With passing time, the active region further evolves, and more and more white pixels appear. The ratio maps also show that the pixels with lower ratios are present all through the evolution of the active region. However, during later stages of the development, the white pixels representing ratios closer to two are dominant and the pixels with ratios less than two were observed primarily at the peripheral regions (similar to that reported by \cite{resonant_sct}). 

We further note that there are significant number of pixels showing the intensity ratios higher than 2. Such observations have also been reported by \cite{resonant_sct} and are attributed to resonant scattering. We note that in the present study, we find that $\sim$~20\% of the pixels show \ion{Si}{4} ratios significantly greater than two. This is significantly larger than those obtained by \cite{resonant_sct}, which is 2.4\%. This may be attributed to the fact that the active region under study is an emerging active region. The distribution shown in Fig.~\ref{fig:fig7} shows that the number of pixels with ratios larger than 2 are present throughout the course of evolution of the active region. In Fig.~\ref{fig:fig6}, we locate the pixels with ratios larger than 2 (panel b), 2.1 (panel c), 2.2 (panel d), 2.3 (panel e) and 2.4 (panel f) above the intensity map. The figure reveals that the pixels with ratios larger than 2 are primarily located within the core of the active region when it is fully evolved.

In the solar and spectral data of two \ion{Fe}{17} lines, \cite{RosMM_2008} found that that this ratio was greater than that predicted by theory. In another paper, \cite{keenan_2014} showed the ratio of the two \ion{O}{6} resonance lines can in some instances be greater than two, which is the optically thin value. These authors suggested that this could be explained due to the geometry where the emitters and absorbers are not spatially distinct, and where the geometry is such that resonant pumping of the upper level has a greater effect on the observed line intensity than resonant absorption in the line-of-sight. \cite{raymond_2000} suggested an alternative explanation, i.e. that the brightness of a low-density region next to a bright could be significantly enhanced by scattered photons. 

In order to check if such an explanation is possible for the results obtained here, in Fig.~\ref{fig:distance} we show scatter plot for intensity ratio (R$>$2 in panel a and R$<2$ in panel b) as a function of distance to the nearest bright point. The bright point is defined as those with intensities larger than 3000~ergs~cm$^{-2}$~s$^{-1}$~sr$^{-1}$. Note that we have only considered the pixels with statistically significant deviation in line ratio from optically thin value (see Section~3.4). The corresponding histograms are shown in panels c and d. Our analysis shows that $\sim$78\% ($\sim$79\%) of pixels with ratio greater (lesser) than 2 are close to a bright region. In such a scenario, the locations with R$>$2.0 may be explained in a manner similar to that by \cite{raymond_2000}. The remainder can be interpreted via the mechanism suggested by \cite{RosMM_2008, keenan_2014}. However, the locations with R$<$2.0 requires further detailed modelling.

\subsection{Distribution of the Intensity Ratios}
\begin{figure}[h]
\centering
\includegraphics[width=0.8\textwidth]{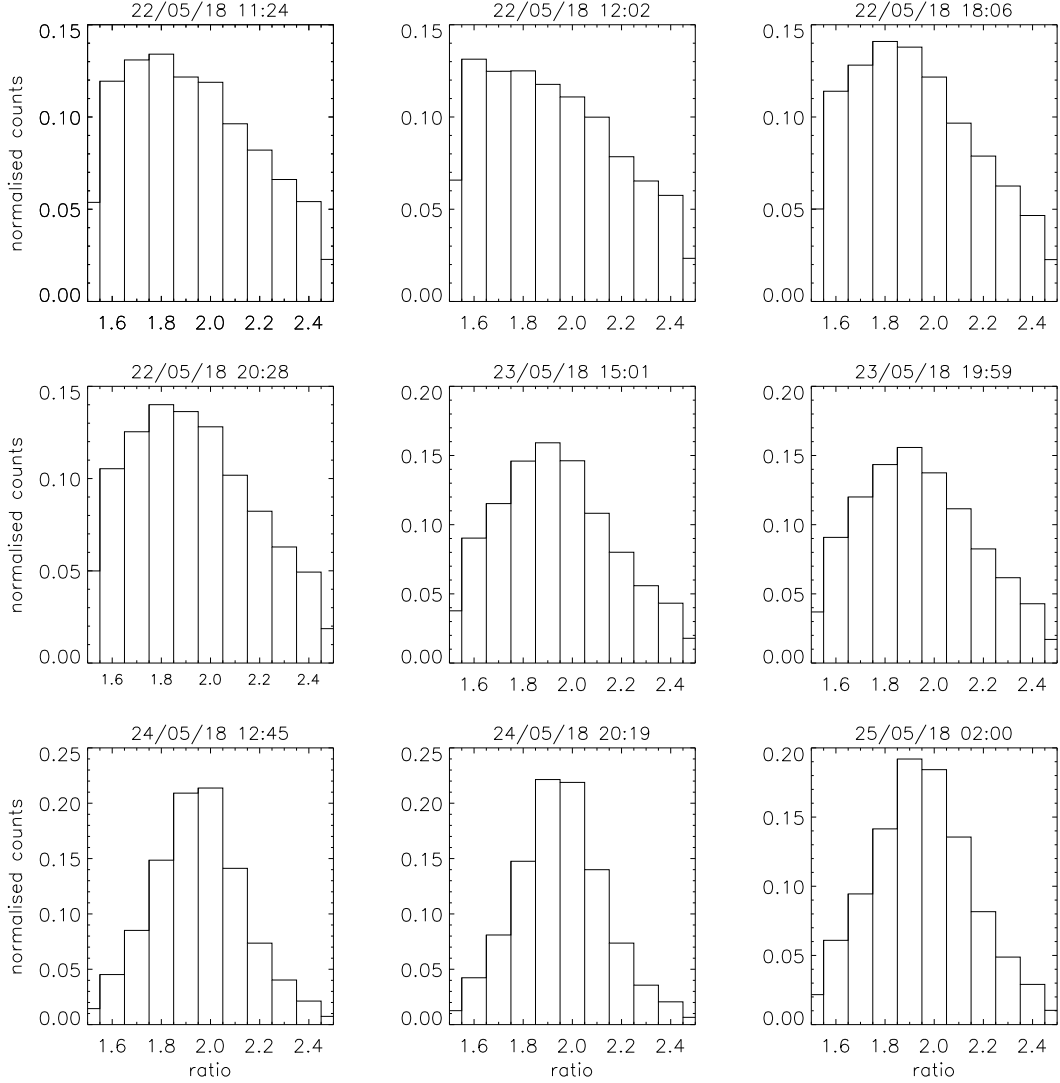}
\caption{Evolution of the distribution of intensity ratios.}
\label{fig:fig7}
\end{figure}

To understand the statistical behaviour, we plot the histograms of the intensity ratios for all 31 rasters. A subset of those depicting different phase of evolution is shown in Fig.~\ref{fig:fig7}. Note that all the distributions are plotted with precisely the same bin size of 0.1. The histograms demonstrate a clear pattern in the distribution. During the initial stages of flux emergence, histograms are asymmetric, and they become more symmetric as the active region evolve. The peaks are close to 1.7 and 1.8 during the initial stage and remain constant at 1.9 during the growth phase. There are significantly large number of pixels showing ratios smaller than 2. We further note that, though smaller, there are significant number of pixels showing ratios larger than 2. With the evolution, the number of pixels showing both smaller as well as larger intensity ratios reduces and become minimal at the time of full evolution of the active region.

\begin{figure}[h]
 \centering
 \includegraphics[width=0.8\textwidth]{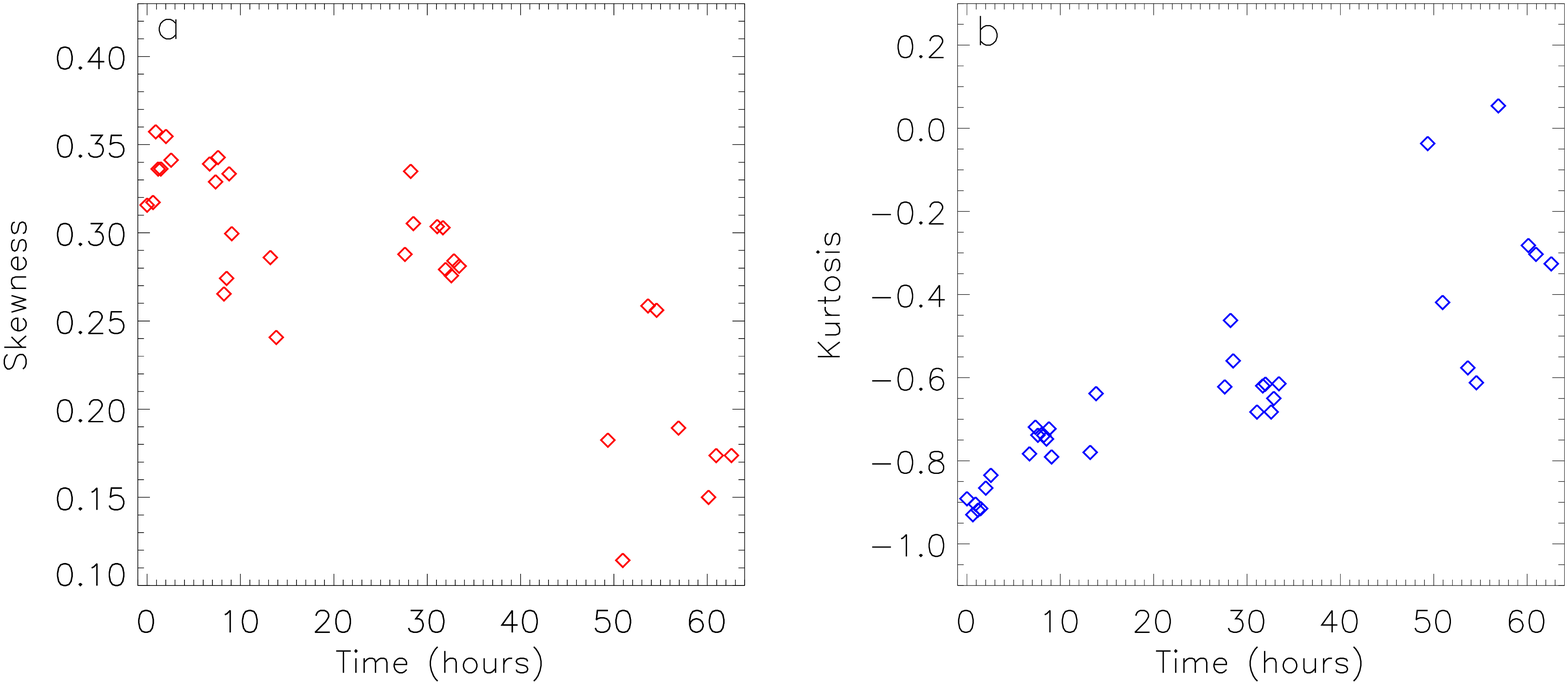}
 \caption{Temporal evolution of skewness (panel a) and kurtosis (panel b) of the distribution of intensity ratio distribution.} \label{fig:fig9}
\end{figure}
\begin{figure}[h]
 \centering
 \includegraphics[width=0.8\textwidth]{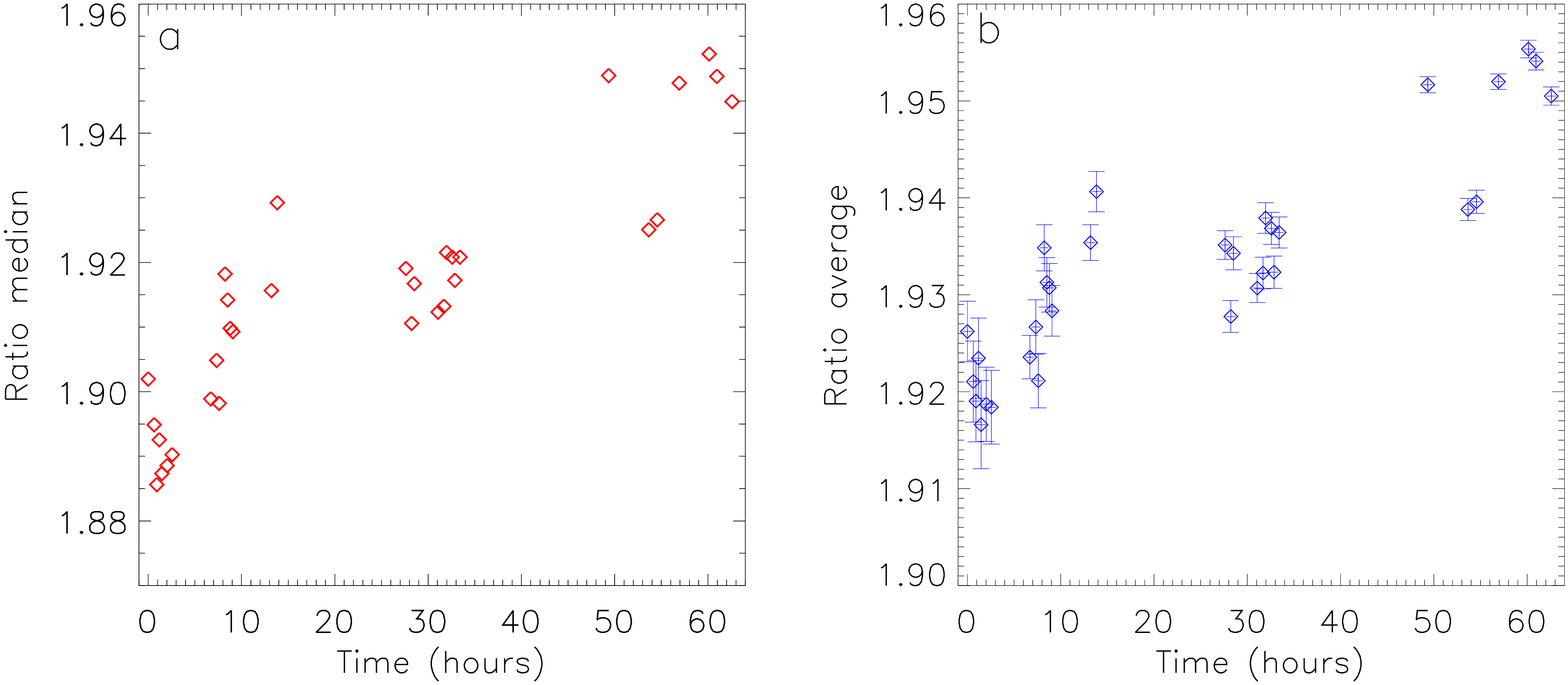}
 \caption{Temporal evolution of the median and mean of the distribution of intensity ratios.}
 \label{fig:fig8}
\end{figure}

In order to confirm the asymmetry of the ratio distribution during the initial stage of flux emergence, we calculated skewness and the kurtosis of the ratio distribution in each raster and plotted them as a function of time in Fig.~\ref{fig:fig9}. The skewness, which is a measure of asymmetry in the distribution, is shown in panel (a), whereas the kurtosis, which is the measure of the flatness of the distribution is plotted in panel (b). As can be noted, the skewness is more substantial at the beginning stage, and it decreases as the active region grow indicating that the distribution is asymmetric during the initial phase of flux emergence and it becomes symmetric during the evolution of the active region. The kurtosis plot further confirms that the distribution is more peaked during the later phase of the evolution when the active region is older. 

In Fig.~\ref{fig:fig8}, we plot the median (left panel) and mean (right panel) of the ratio distribution from all 31 observations as a function of time. In each plot, time zero represents the initial observation time 11:24 UT on May 22 and the difference in time is calculated by taking this observation time as the reference. The left panel of Fig.~\ref{fig:fig8} demonstrate a clear trend of increasing median with time. During the early stage of the flux emergence, the median of the ratio distribution is lower than 2 that increases with the evolution of the active region. Similarly, the mean of the ratio distribution increases with time (right panel of Fig.~\ref{fig:fig8}. The larger error bar in initial observations is due to a smaller number of data points. The gradual increase in median and mean of the ratio distribution is suggestive of the fact that opacity effects are much larger during the early stages of flux emergence and decreases with the age of the active region.

\subsection{Intensity Ratios in Quiet Sun}
\begin{figure}[h]
 \centering
 \includegraphics[width=0.8\textwidth]{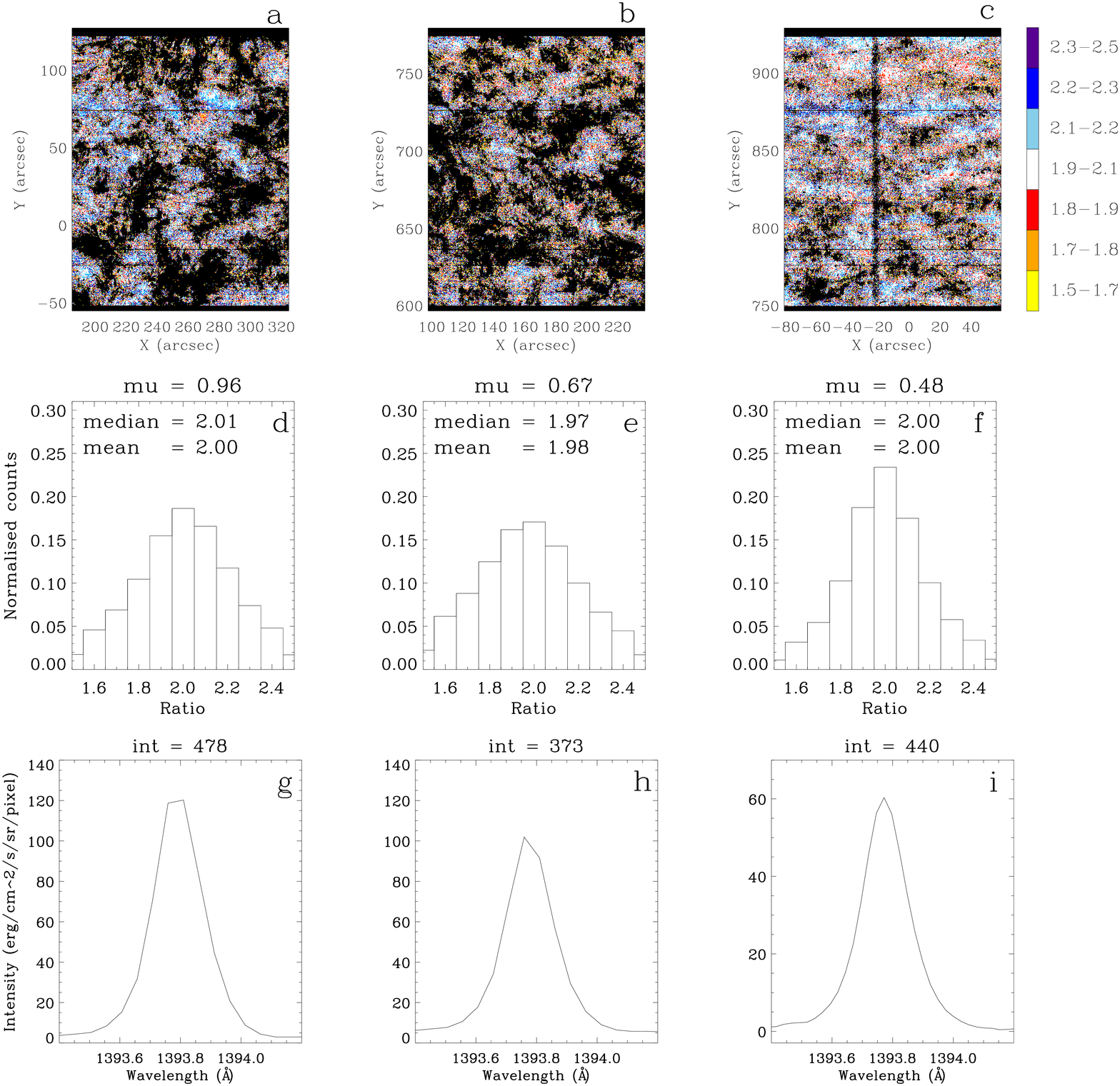}
 \caption{Top panel (a,b,c): QS ratio maps obtained using the observation from different location on the disk of the Sun. Middle panel (d,e,f): Corresponding ratio histograms. Bottom panel (g,h,i): Average spectra of the raster.}
 \label{fig:fig10}
\end{figure}

It is a known fact that the intensities of the optically thick spectral lines show centre-to-limb-variation (CLV). Since the 31 IRIS rasters studied here are observed at different spatial locations on the Sun, the CLV may affect our results if \ion{Si}{4} lines are optically thick in all kinds of spatial structures on the Sun. Therefore, we turn our attention to Quiet Sun observations. For this purpose, we have studied three different IRIS rasters recorded at different locations on the Sun to obtain the intensity ratios of the \ion{Si}{4} lines. We note that we have applied the same criteria for quiet Sun spectra as detailed in \S\ref{sec:od}. 

We plot the obtained ratio maps and the corresponding distributions for the three quiet Sun rasters in Fig.~\ref{fig:fig10}. The colour scheme shown here is exactly the same as for the active regions, shown in Fig.~\ref{fig:fig4}. The mean and median of the distribution is also labelled. The plots clearly show that the ratio maps are dominated by white pixels, showing the ratios lying between 1.9 and 2.1. The distribution of the ratios looks symmetric with a peak at 2.0 for all the three rasters, irrespective of their spatial location. For both limb and disk centre observations the median of the ratios are very close to 2. The spectra shown in the bottom panel of Fig.~\ref{fig:fig10} also confirm that the intensity in the Quiet Sun is not changing as a function of location. 

\begin{figure}[h]
 \centering
 \includegraphics[width=0.9\textwidth]{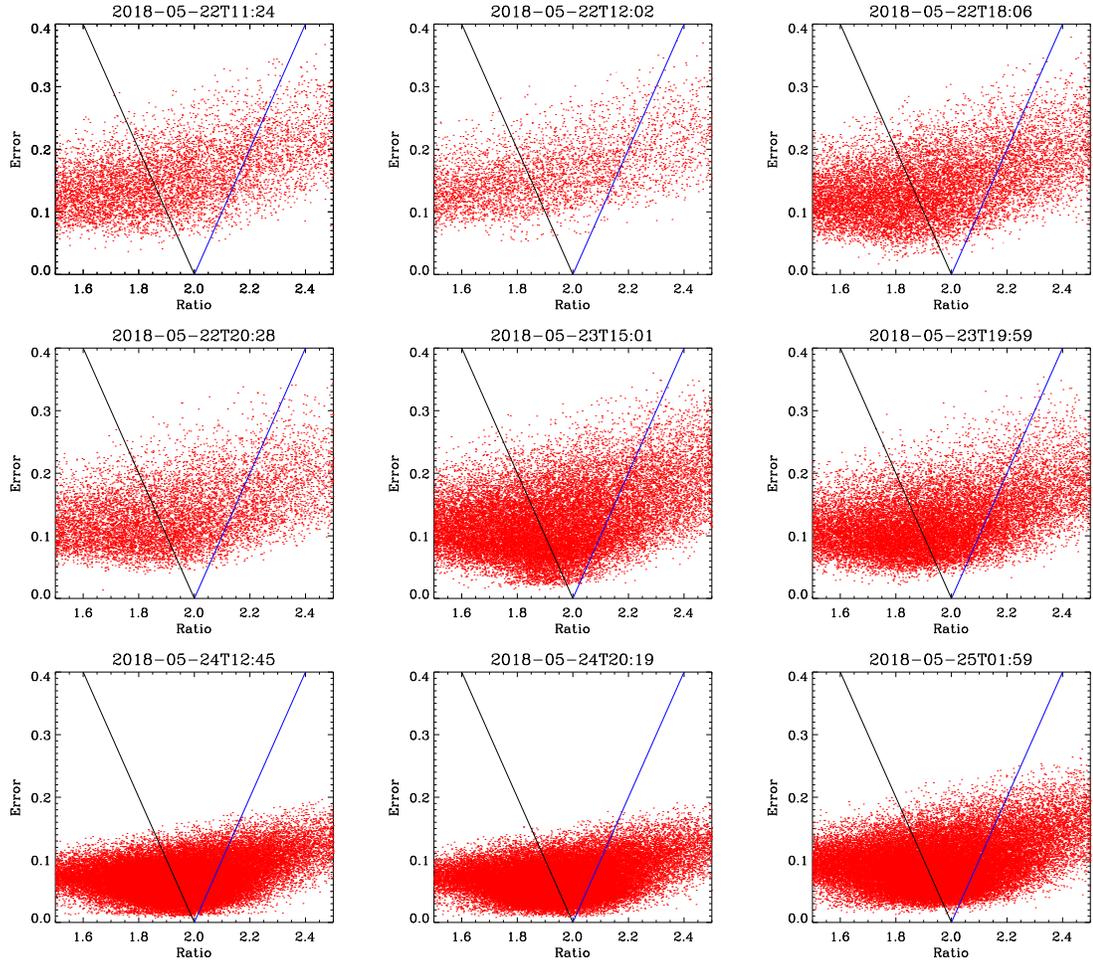}
 \caption{Error in \ion{Si}{4} line ratio as a function of line ratio for nine different observations taken during the evolution of the active region.}
 \label{fig:fig11}
\end{figure}
\begin{figure}[h]
 \centering
 \includegraphics[width=0.9\textwidth]{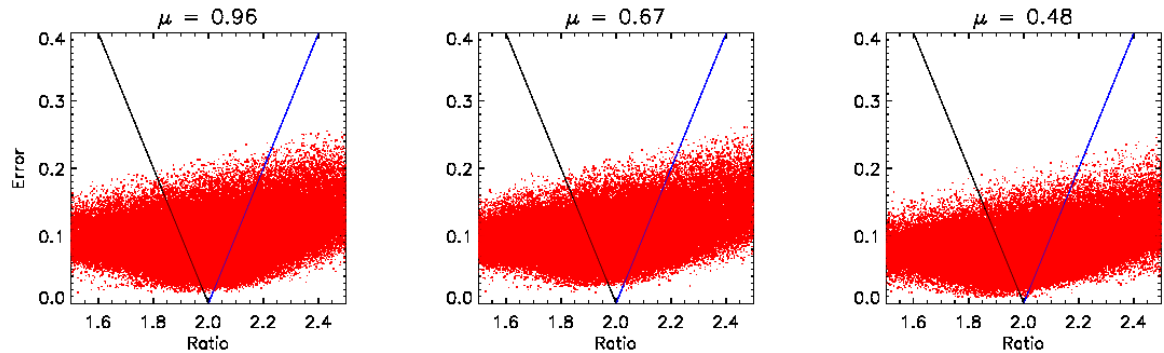}
 \caption{Error in \ion{Si}{4} line ratio as a function of line ratio for the quiet Sun rasters observed at different $\mu$ values.}
 \label{fig:fig12}
\end{figure}
\begin{figure}[h]
 \centering
 \includegraphics[width=\textwidth]{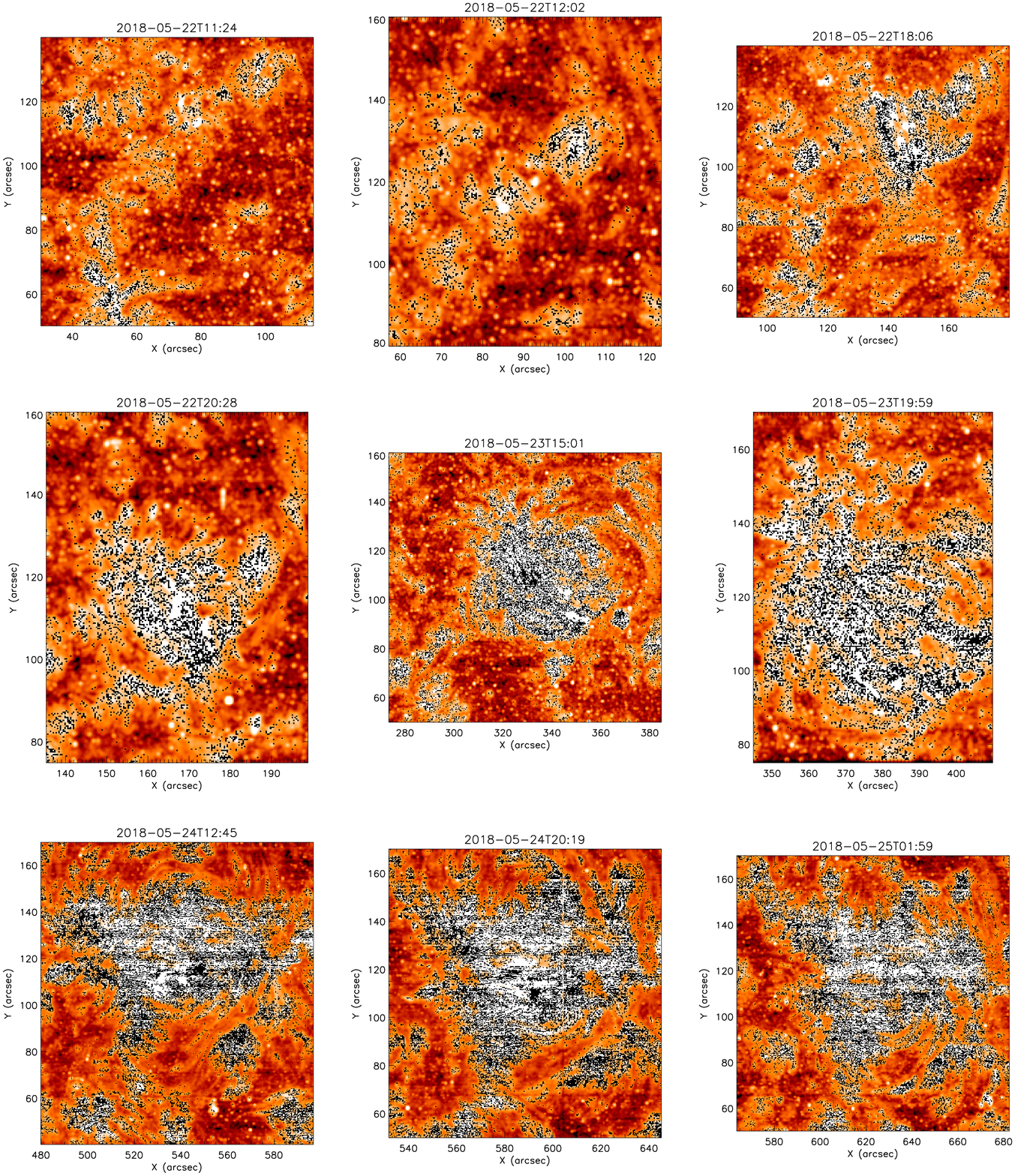}
 \caption{Location of pixels with significantly reduced line ratio ( $r+\delta r < 2$) during the evolution of active region. Pixels with $r+\delta r < 2$ are colour coded in black on \ion{Si}{4} 1394 {\AA} intensity map}
 \label{fig:fig13}
\end{figure}
\begin{figure}[h]
 \centering
 \includegraphics[width=\textwidth]{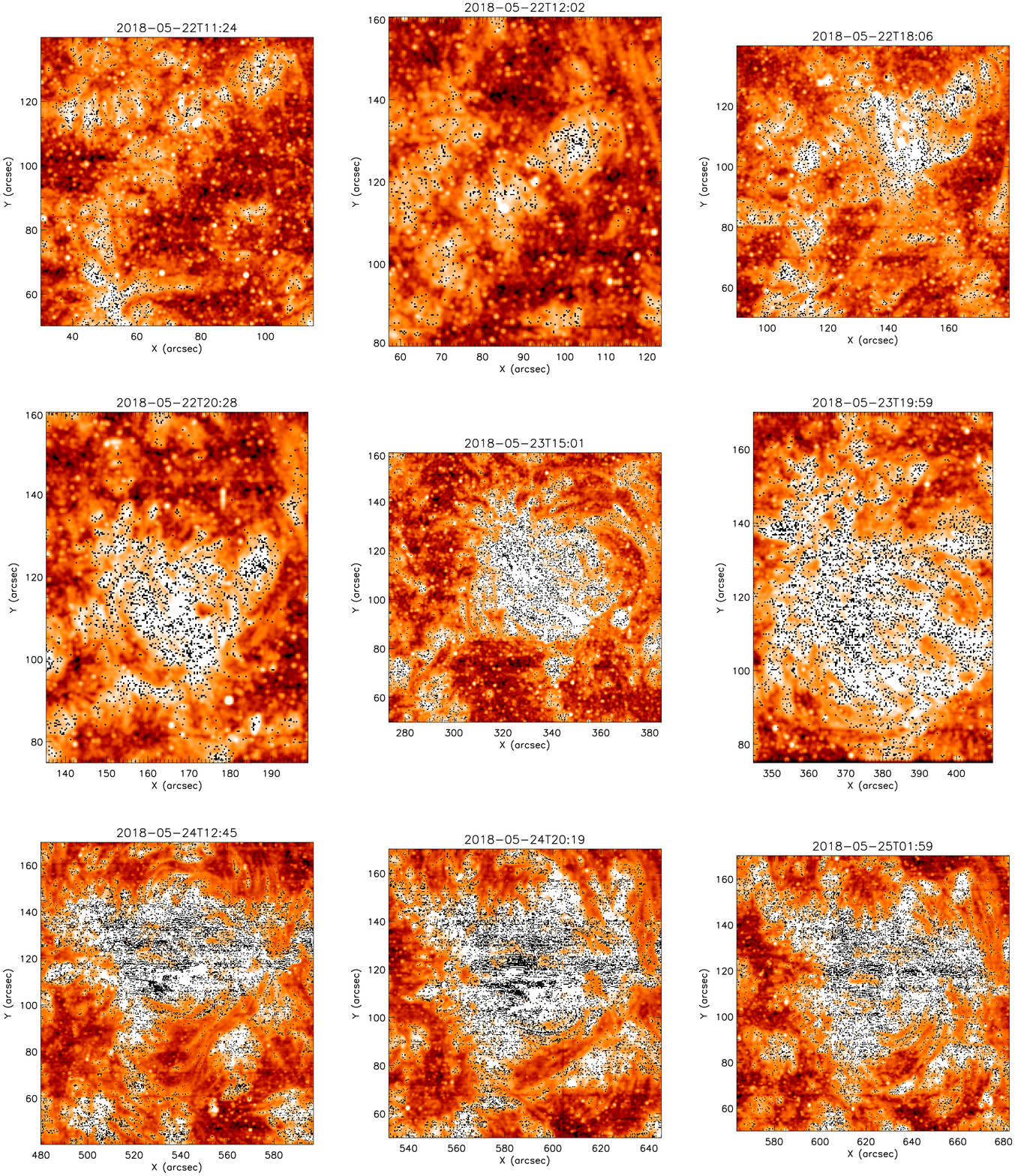}
\caption{Location of pixels with significant enhancement in line ratio ( $r-\delta r > 2$) during the evolution of active region. Pixels with $r-\delta r > 2$ are colour coded in black on \ion{Si}{4} 1394 {\AA} intensity map} \label{fig:fig14}
\end{figure}
\subsection{Error in \ion{Si}{4} line ratio}\label{error}
Here we calculate the uncertainty in the observed line ratio to identify the pixels which show statistically significant deviation from the theoretical line ratio. Assuming the Poisson error statistics for photons, we first calculated the error in integrated line intensities of \ion{Si}{4} lines and then propagated the errors to obtain the uncertainty in line ratios (see Appendix~\ref{error} for details). In Fig.~\ref{fig:fig11} we plot the error in line ratio ($\delta r$) as a function of line ratio (r) for nine different observations, spread over the duration of the flux emergence. On each plot, line $r-\delta r = 2$ is shown in blue colour and the line $r+\delta r = 2$ is plotted in black colour. These lines are plotted to identify the pixels with statistically significant deviation. Pixels on the left side of the black line have ratio plus the measurement error on the ratio $r + \delta r <2 $. On the other hand pixels on the right side of the blue line have $r-\delta r>2 $. Therefore these pixels can be considered as the pixels with significant deviation in line ratio. There are a large fraction of pixels on the left side of the black line, which indicates the opacity effects in the active region. Similar plots for the quiet Sun are shown in Fig.~\ref{fig:fig12}. From Fig.~\ref{fig:fig12} it is clear that pixels with a statistically significant ratio above and below 2.0 are equally present in the Quiet Sun.

The spatial location of pixels with a statistically significant reduction in line ratios are shown in Fig~\ref{fig:fig13}. Here we display these pixels on \ion{Si}{4} 1394 {\AA} intensity map which are observed during the flux emergence. Location of pixels with significantly reduced line ratios are colour coded in black. From Fig~\ref{fig:fig13} it is clear that most of the opacity affected pixels are located at the periphery of the active region. Similarly, the location of pixels with statistically significant ratio above 2 are shown in Fig~\ref{fig:fig14}. These pixels are densely populated close to the active region core.
   
\section{DISCUSSION AND CONCLUSIONS}\label{sec:conclusion}
  
 In this paper, we study the temporal evolution of \ion{Si}{4} line ratio in an EFR. The obtained results are summarised as follows. Throughout the development of the EFR, the histograms of the ratio are asymmetric (skewed as well as shifted) towards the low ratio values. This tendency is stronger during the initial stage of flux emergence. Both the average and median of the distribution were lower than 2 in the initial phase, and they gradually approached to 2, which is the theoretically expected value for optically thin plasmas. On the other hand, in the quiet Sun, the distribution is highly symmetric and peaks at 2. There are also significant number of pixels with ratios larger than 2. Our analysis shows that the pixels with ratios smaller than 2 are predominantly located at the periphery of the active region whereas those with ratios larger than 2 are in the core. 
  
The reduced \ion{Si}{4} ratio is usually attributed to optical effect, as the line with the largest oscillator strength is strongly affected by opacity. In EFRs, particularly during the early phase, dense chromospheric plasma is lifted to the corona by emerging loops. Heating of the plasma often occurs intermittently due to the filamentary structure in the rising magnetic flux \citep{isobe}. The magnetic reconnection between the neighbouring rising loops also causes small scale brightenings in and above the chromospheric heights \citep[][]{IsoTA, peter, GupT}. Such dynamic events increase the densities along the line of sight resulting into reabsorption of photons by \ion{Si}{4} ions. Since the probability of reabsorption for the1394 line is a factor 2 greater than for the 1403 line, so its intensity gets reduced more and the ratios less than 2 is observed.

Our result showed that the fraction of the pixels with \ion{Si}{4} ratio less than 2 is larger in the emerging active region compared with the quiet sun. Moreover, the distribution is more skewed and shifted toward lower value in the early phase of the emergence than in the later phase. Also, from the measured line ratio error, it is clear that the large fraction of active region pixels shows a statistically significant reduction in line ratio. These results support the interpretation that larger opacity may be the primary reason for the smaller values of the \ion{Si}{4} ratio. 

A correlation study of pixels with ratios smaller (larger) than 2 shows that about ~79\% (~78\%) of pixels are close to a bright region ($>$3000~ergs~cm$^{-2}$~s$^{-1}$~sr$^{-1}$). The results for R$>$2 is in the line of suggestion made by \cite{raymond_2000} that the brightness of a low-density region next to a bright could be significantly enhanced by scattered photons. However, the dichotomy persists for locations with R$<$ 2 and further study in \ion{Si}{4} modelling is required to understand the \ion{Si}{4} emission in the transition region in greater detail.

Finally, the observation that the \ion{Si}{4} line intensity ratios are significantly smaller than two and progressively changes to 2 with the evolution of the active region may hold potential diagnostics for the understanding of the physics of flux emergence. Such an analysis will require full-scale magneto-hydrodynamic simulations of emerging flux regions and forward modelling to shed further lights, which is out of the scope of this paper.

\begin{acknowledgements}
We thank an anonymous referee for the insightful comments that has helped improve the paper. DT thanks J. A. Klimchuk for initial discussions on this subject. DT and NVN acknowledged the support from the Max-Planck India Partner Group of MPS at IUCAA. HI is supported by JSPS KAKENHI Grant Number 18H01254. Armagh Observatory and Planetarium is core funded by the Northern Ireland Government through the Department for Communities. IRIS is a NASA small explorer mission developed and operated by LMSAL with mission operations executed at NASA Ames Research center and major contributions to downlink communications funded by ESA and the Norwegian Space Centre. We would like to thank AIA, HMI, and IRIS teams for providing valuable data. CHIANTI is a collaborative project involving George Mason University, the University of Michigan (USA), University of Cambridge (UK) and NASA Goddard Space Flight Center (USA). NVN is funded via a studentship from Armagh Observatory and Planetarium.
\end{acknowledgements}
\bibliographystyle{apj}
\bibliography{ref}
\appendix
\section{Error Analysis}\label{error}
The ratio of \ion{Si}{4} lines are defined as
\begin{align}
R &=\frac{I_{1394}}{I_{1403}}
\end{align}
Where $I_{1394}$ and $I_{1403}$ are the integrated line intensities of 1394 and 1403 {\AA} respectively.  
Error in the line ratio can be obtained using the formula
\begin{align}
\frac{\delta R}{R} &=\sqrt{\left(\frac{\delta I_{1394}}{I_{1394}}\right)^{2}+\left(\frac{\delta I_{1403}}{I_{1403}}\right)^{2}}
\end{align}

Where $\delta R$ is the line ratio error. $\delta I_{1394}$ and $\delta I_{1403}$ are the error integrated line intensities of 1394 and 1403 {\AA} respectively.
Error in the intensity measurement can be calculated using photon statistics. To do this first we have to convert the intensities in DN unit to the photon counts. This can be done using the formula
\begin{align}
I(photons) &=\frac{g}{yield}\times I(DN)
\end{align}
Where $g$ is the gain which is the number of electrons released in the detector that yield 1 DN. Yield is the number of electrons released by one incident photon. For FUV spectra gain in 6 and the yield is 1.5 \citep{iris}.
Therefore the above equation can be rewritten as
\begin{align}
I(photons) &=4\times I(DN)
\end{align}
According to the photon statistics, the error in photon counts is the given by,
\begin{align}
\delta I(photons) &=\sqrt{I(photons)}
\end{align}
Using the above equations, error in the line ratio can be written as
 \begin{align}
\delta R &=R \times \sqrt{\frac{1}{I_{1394}(photons)}+\frac{1}{I_{1403}(photons)}}
\end{align}
\section{Details of the Active Region and Quiet Sun Observations}
\begin{table}
\centering
\begin{tabular}{ |c|c|c|c|c|c| }
\hline
Data  & Date of      & Time of      & FOV   & Exposure & $\mu$-value \\
    & Observation    & Observation (UT) & (arcsec) & time (S) &   \\
\hline
Set 1 & 22-May-2018 & 11:24:59 & $112''\,\times175\,''$ & 4 & 0.99 \\
Set 2 & 22-May-2018 & 12:02:33 & $67''\,\times119\,''$ & 4 & 0.98 \\
Set 3 & 22-May-2018 & 12:19:11 & $67''\,\times119\,''$ & 4 & 0.98 \\
Set 4 & 22-May-2018 & 12:35:48 & $67''\,\times119\,''$ & 4 & 0.98 \\
Set 5 & 22-May-2018 & 13:25:40 & $67''\,\times119\,''$ & 4 & 0.98 \\
Set 6 & 22-May-2018 & 13:58:55 & $67''\,\times119\,''$ & 4 & 0.98 \\
Set 7 & 22-May-2018 & 18:06:37 & $112''\,\times175\,''$ & 4 & 0.98 \\
Set 8 & 22-May-2018 & 18:44:11 & $67''\,\times119\,''$ & 4 & 0.98 \\
Set 9 & 22-May-2018 & 19:00:49 & $67''\,\times119\,''$ & 4 & 0.98 \\
Set 10 & 22-May-2018 & 19:38:30 & $67''\,\times119\,''$ & 4 & 0.98 \\
Set 11 & 22-May-2018 & 19:55:08 & $67''\,\times119\,''$ & 4 & 0.98 \\
Set 12 & 22-May-2018 & 20:11:45 & $67''\,\times119\,''$ & 4 & 0.97 \\
Set 13 & 22-May-2018 & 20:28:23 & $67''\,\times119\,''$ & 4 & 0.97 \\
Set 14 & 23-May-2018 & 00:36:37 & $112''\,\times175\,''$ & 4 & 0.97 \\
Set 15 & 23-May-2018 & 01:14:11 & $67''\,\times119\,''$ & 4 & 0.97 \\
Set 16 & 23-May-2018 & 15:01:20 & $112''\,\times175\,''$ & 4 & 0.93 \\
Set 17 & 23-May-2018 & 15:38:54 & $67''\,\times119\,''$ & 4 & 0.92 \\
Set 18 & 23-May-2018 & 15:55:32 & $67''\,\times119\,''$ & 4 & 0.92 \\
Set 19 & 23-May-2018 & 18:28:17 & $112''\,\times175\,''$ & 4 & 0.91 \\
Set 20 & 23-May-2018 & 19:05:51 & $67''\,\times119\,''$ & 4 & 0.91 \\
Set 21 & 23-May-2018 & 19:22:29 & $67''\,\times119\,''$ & 4 & 0.91 \\
Set 22 & 23-May-2018 & 19:59:30 & $67''\,\times119\,''$ & 4 & 0.91 \\
Set 23 & 23-May-2018 & 20:16:08 & $67''\,\times119\,''$ & 4 & 0.91 \\
Set 24 & 23-May-2018 & 20:49:23 & $67''\,\times119\,''$ & 4 & 0.90 \\
Set 25 & 24-May-2018 & 12:45:22 & $112''\,\times175\,''$ & 15 & 0.81 \\
Set 26 & 24-May-2018 & 17:03:22 & $112''\,\times175\,''$ & 4 & 0.79 \\
Set 27 & 24-May-2018 & 17:58:37 & $112''\,\times175\,''$ & 4 & 0.79 \\
Set 28 & 24-May-2018 & 20:19:59 & $112''\,\times175\,''$ & 15 & 0.77 \\
Set 29 & 24-May-2018 & 23:32:53 & $112''\,\times175\,''$ & 8 & 0.76 \\
Set 30 & 25-May-2018 & 00:21:51 & $112''\,\times175\,''$ & 8 & 0.75 \\
Set 31 & 25-May-2018 & 01:59:45 & $112''\,\times175\,''$ & 8 & 0.74 \\
\hline
\end{tabular}
\caption{Details of EFR Observations.}
\label{table:tab1}
\end{table} 

\begin{table}[h]
\centering
\begin{tabular}{ |c|c|c|c|c|c| }
\hline
Data  & Date of      & Time of      & FOV   & Exposure & $\mu$-value \\
    & Observation    & Observation (UT) & (arcsec) & time (S) &   \\
\hline
Set 1 & 07-June-2014 & 23:09:36 & $141''\,\times175\,''$ & 15 & 0.67 \\
Set 2 & 07-June-2014 & 12:09:34 & $141''\,\times175\,''$ & 15 & 0.96 \\
Set 3 & 06-March-2014 & 10:04:51 & $141''\,\times175\,''$ & 30 & 0.48 \\
\hline
\end{tabular}
\caption{Details of quiet Sun Observations.}
\label{table:tab2}
\end{table} 
\section{Distribution of \ion{Si}{4} intensity ratios by defining the domain of emerging flux}
The analysis in this paper has been performed for pixels that belonged to the rectangular region surrounding the EFR. Therefore, it is possible that the neighbouring QS pixels that are included in the selected rectangular region could affect our results. To test this, we performed a similar analysis by defining a boundary to the EFR. Photospheric magnetic field strength recorded by HMI on board SDO can be used to define the boundary between EFR and the neighbouring quiet Sun. To do this we co-aligned the LOS magnetogram with IRIS raster using SJI 1400, AIA 1600 images. Then, by manual inspection defined the EFR as the region with absolute magnetic field strength greater than 20 G and performed a similar analysis for the pixels that fall in the EFR region.

\begin{figure}[h]
\centering
\includegraphics[width=0.8\textwidth]{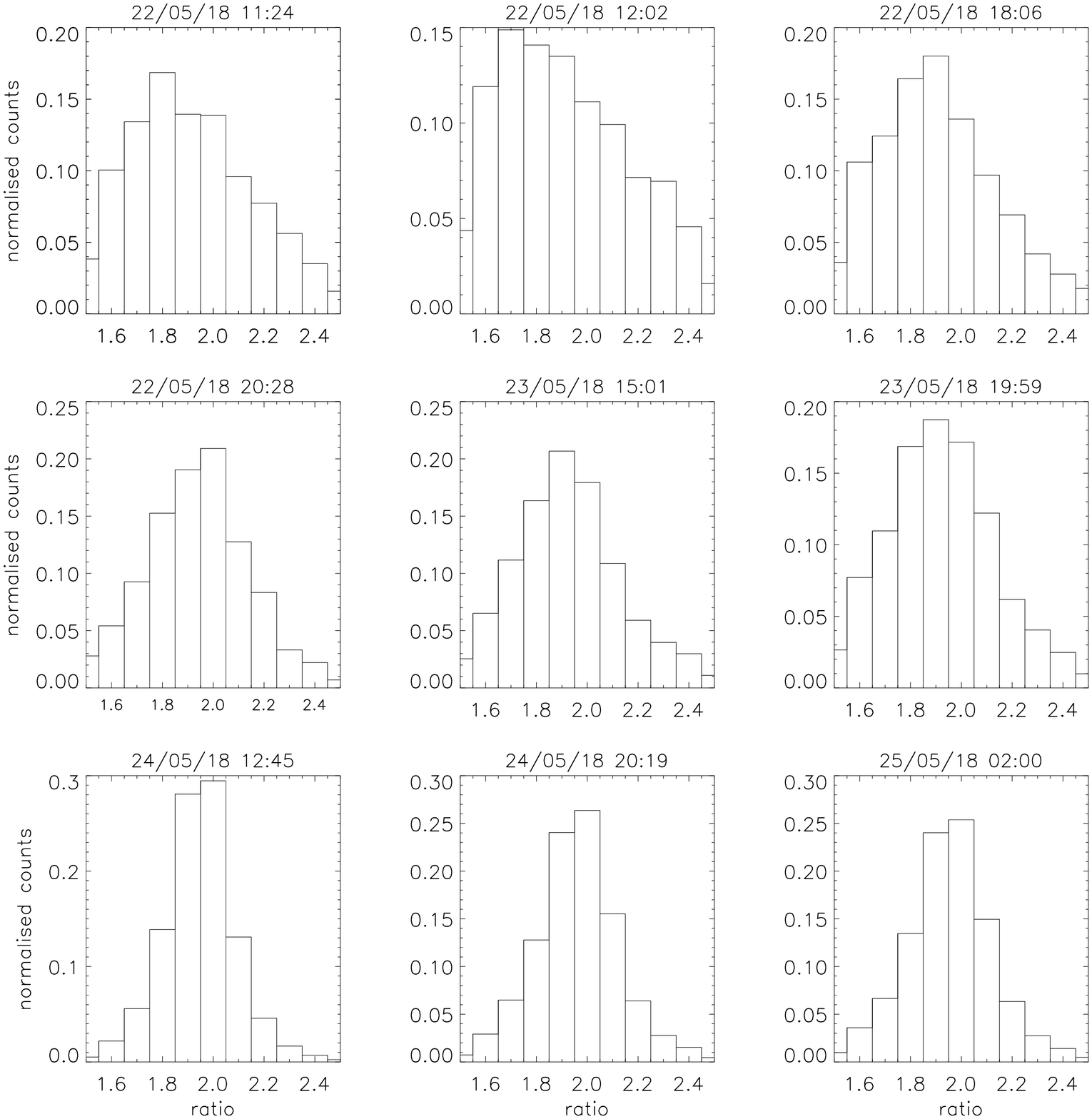}
\caption{Evolution of the distribution of intensity ratios.}
\label{fig:figc1}
\end{figure}

\begin{figure}[h]
 \centering
 \includegraphics[width=0.8\textwidth]{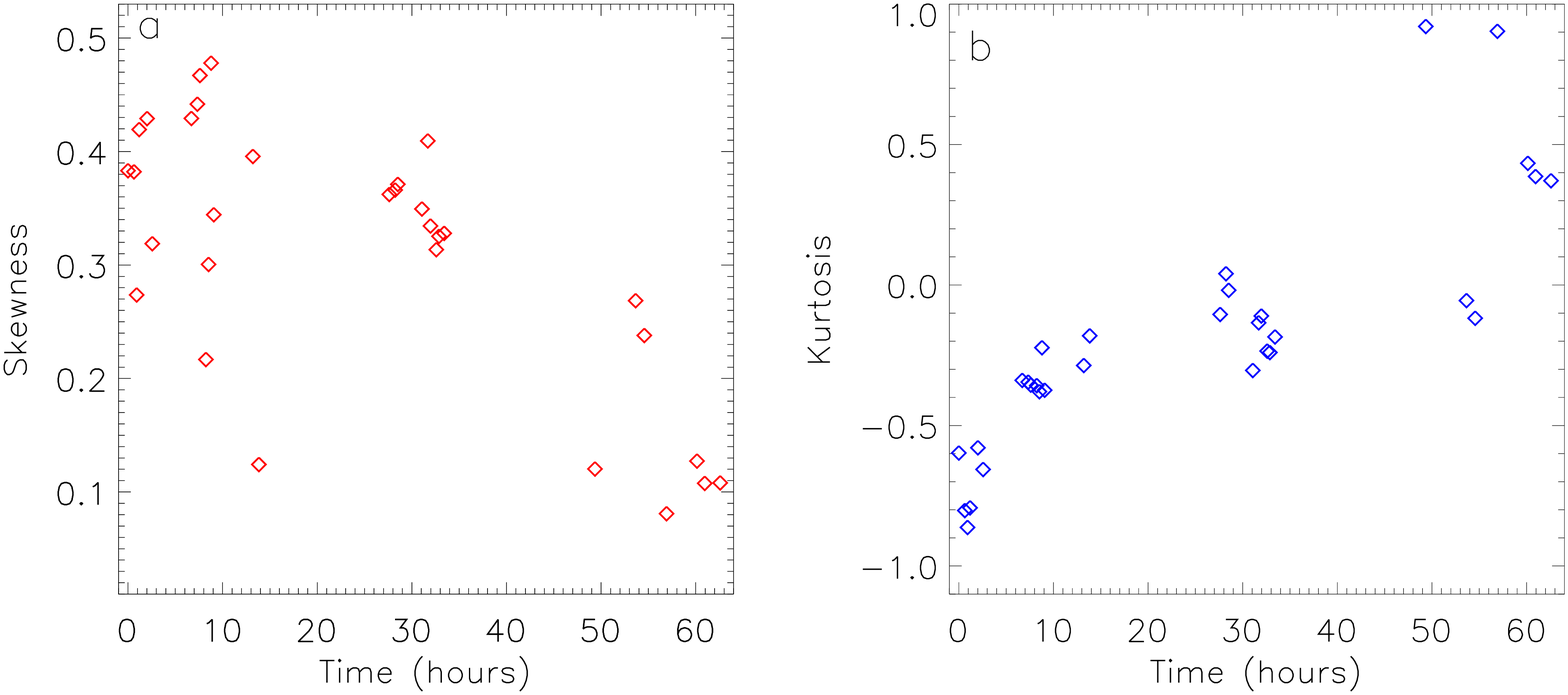}
 \caption{Temporal evolution of skewness (panel a) and kurtosis (panel b) of the distribution of intensity ratio distribution.} \label{fig:figc2}
\end{figure}

The evolution of distribution of \ion{Si}{4} line ratio in EFR is shown in Fig.~\ref{fig:figc1}. Similar to the previous analysis, histograms are asymmetric during the initial stages of flux emergence, and they become more symmetric as the active region evolve. This observation is further confirmed in the skewness plot shown Fig.~\ref{fig:figc2} panel (a), where we see a decrease in the skewness of the ratio distribution as the active region evolve. The peaks ratios in Fig.~\ref{fig:figc1} are also close to 1.7 and 1.8 during the initial stage and remain constant at 1.9 during the growth phase as seen in the previous analysis. The kurtosis plot shown in panel (b) of Fig.~\ref{fig:figc2} confirms that the distribution is more peaked during the later phase of the evolution when the active region is older.

\begin{figure}[h]
 \centering
 \includegraphics[width=0.8\textwidth]{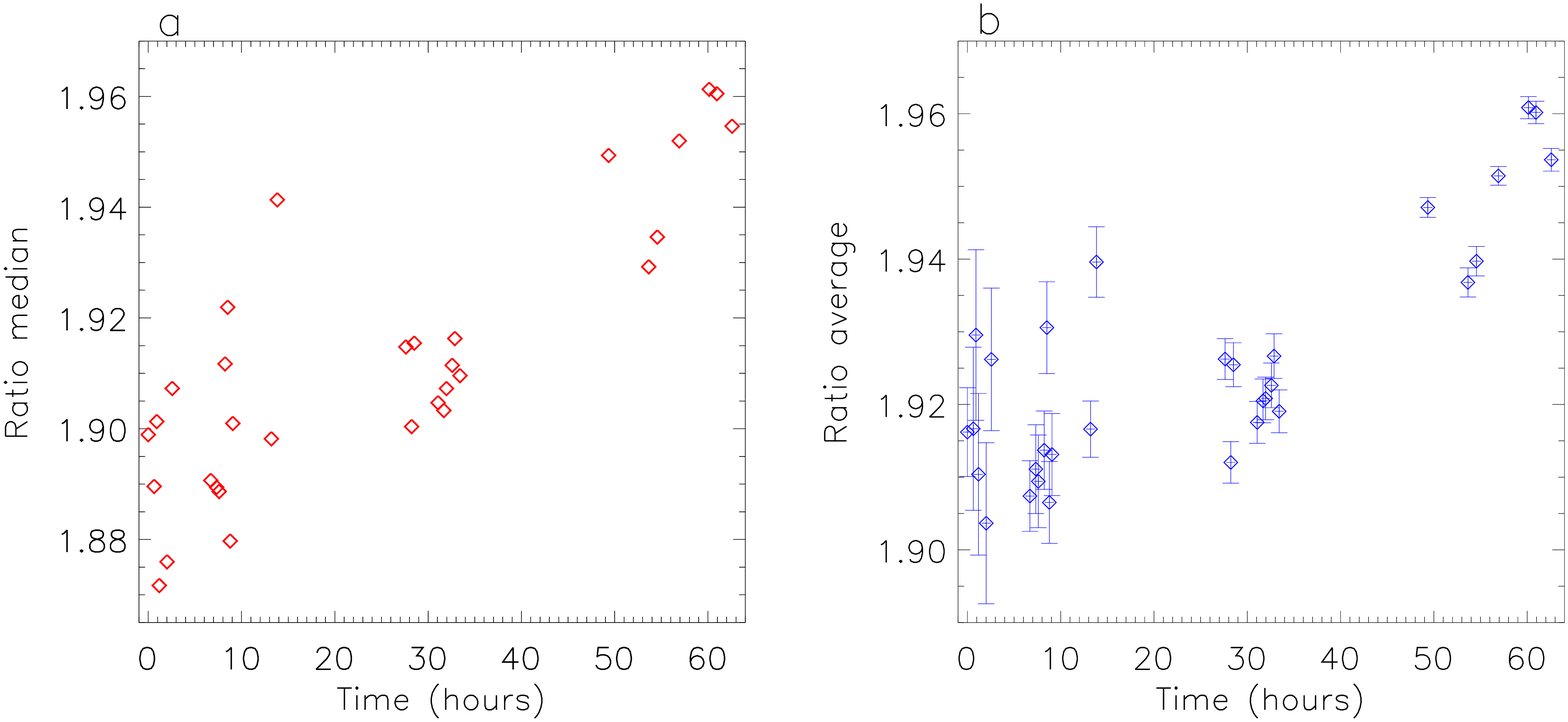}
 \caption{Temporal evolution of the median and mean of the distribution of intensity ratios.}
 \label{fig:figc3}
\end{figure}

Finally, we plot the evolution of median and average values of intensity ratios in Fig.~\ref{fig:figc3}. From the figure, it is clear that both the median and the average value of intensity ratios are smaller than 2 during the initial stage of flux emergence and they both gradually show an increase as the active region evolves. Large error bars in the average line at the beginning are due to the less number of data points with an absolute magnetic field strength greater than 20 Gauss. The above results confirm that the opacity effects are stronger during the initial stage of flux emergence and decreases as the active region evolve. Therefore, we conclude that results obtained by defining a rectangular boundary to the EFR and by a boundary based on magnetic field strength are very similar. 


\end{document}